\documentclass[10pt,USenglish,twocolumn]{article}\usepackage[top=.56in,left=.7in,right=.7in,bottom=.99in]{geometry}
\usepackage{graphicx}
\usepackage{amssymb}
\usepackage{babel}
\usepackage{amsmath}
\usepackage{natbib}
\usepackage{tgtermes}
\usepackage[mathcal]{euscript}
\usepackage{enumerate}
\usepackage[utf8]{inputenc}
\usepackage[T1]{fontenc}
\usepackage{epstopdf}
\usepackage{setspace}
\title{Algorithms and Epistemic Diversity / Cognitive Biases}
\title{Algorithmic Distortion of Cognitive Processes}
\title{Algorithmic Distortion of Informational Landscapes
}

\author{Camille Roth\\\small \href{http://cmb.huma-num.fr}{Computational Social Science Team}\vspace{-.5em}\\\small Centre Marc Bloch, CNRS/HU Berlin\vspace{-.5em}\\\small Friedrichstrasse 191, D-10117 Berlin\vspace{-.5em}\\\small roth@cmb.hu-berlin.de\vspace{-.5em}}
\date{%
\footnotesize\bf (to appear in 2019 in {\em Intellectica, vol. 70)}
}
\usepackage[usenames,dvipsnames]{color}
\usepackage[bookmarks, colorlinks, breaklinks, pdftitle={},pdfauthor={Camille Roth}]{hyperref}
\hypersetup{linkcolor= MidnightBlue,citecolor= MidnightBlue,filecolor=black,urlcolor= MidnightBlue}
\definecolor{gray}{rgb}{0.98,.98,.98}
\definecolor{darkgray}{rgb}{0.8,.8,.8}
\definecolor{darkdarkgray}{rgb}{0.66,.66,.66}

\newcommand{\remove}[1]{}

\newcommand{\tb}[1]{\textcolor{blue}{#1}}

\newcommand{\itx}[1]{\begin{itemize}#1\end{itemize}}

\newcommand{\x}{\item}

\makeatletter
\renewcommand{\section}{\@startsection{section}{1}{0mm}
{\baselineskip}{\baselineskip}{\raggedright\Large\bf\textcolor{MidnightBlue}}}
\makeatother

\begin{document}
\maketitle
\begin{abstract}
\noindent The possible impact of algorithmic recommendation on the autonomy and free choice of Internet users is being increasingly discussed, especially in terms of the rendering of information and the structuring of interactions. This paper aims at reviewing and framing this issue along a double dichotomy. The first one addresses the discrepancy between users’ intentions and actions (1) under some algorithmic influence and (2) without it. The second one distinguishes algorithmic biases on (1) prior information rearrangement and (2) posterior information arrangement. In all cases, we focus on and differentiate situations where algorithms empirically appear to expand the cognitive and social horizon of users, from those where they seem to limit that horizon. We additionally suggest that these biases may be best appraised by taking into account the underlying social processes which algorithms are building upon.
\medskip

\bigskip\vfill

%
\noindent {\em \footnotesize L'impact potentiel de la recommandation algorithmique sur le libre-arbitre des internautes fait l'objet d'un débat de plus en plus visible.  Ces discussions sont notamment centrées sur le biais induit par les algorithmes dans le traitement et la structuration à la fois de l'information et des interactions en ligne. Cet article vise à évoquer et organiser l'état de l'art correspondant suivant une double dichotomie. La première concerne l'écart entre les intentions et actions des utilisateurs (1) en l'absence d'algorithmes et (2) en faisant usage, d'une manière ou d'une autre, d'algorithmes de recommandation. La seconde distingue l'influence des algorithmes (1) sur la réorganisation de l'information (a priori) et (2) sur sa restitution aux utilisateurs (a posteriori). Dans tous les cas, nous nous attachons à distinguer les situations où l'on peut montrer empiriquement que les algorithmes tendent à ouvrir ou bien au contraire limiter l'horizon cognitif des utilisateurs. Nous insistons en outre sur le caractère intrinsèquement social de la recommandation algorithmique, dont l'effet ne peut être entièrement compris et évalué sans prendre en compte les dynamiques collectives sous-jacentes au calcul implémenté par ces algorithmes.
\medskip

\noindent {\textbf{Mots-clés:}}
algorithmes; recommandation algorithmique; biais cognitifs; bulle de filtre; diversité informationnelle.
}
\end{abstract}



\section*{Introduction}
The algorithms ruling online platforms and mediating many of our digital interactions have been at the center of an increasingly visible debate. The main issues revolve around the nature of the potential behavioral distortions that they may engender.
In essence, algorithms often encode and implement a myriad of principles which have long been put into practice by human actors --- think of librarians when navigating large document collections, or opinion leaders who contribute to endorse or validate information in public spaces.

As such, algorithmic devices do not only entail a certain rigidification of many organizational and rational-legal processes (with respect for instance to the regulatory effect of code \citep{lessig2009code}) or a formalization of many social and cognitive processes (for example, each web platform tends to produce an artificial social world governed by a specific set of possible interaction behaviors \citep{marres2017digital}). Their growing ubiquity may also induce a certain number of biases in the processing and structuring of information and interactions, with respect to some baseline human behavior. This paper will essentially focus on the algorithms that populate digital spaces, and more precisely those which are at some level in charge of mediating the access by users to information and which, in turn, may contribute to distort their epistemic landscapes.  We shall question the type and the extent of the cognitive bias which algorithms exert on user actions, on their beliefs and thus on their free will, relatively to an organic reference point (\hbox{i.e.} free of any algorithmic tampering). 

We will principally discuss these issues by developing two dichotomies.
The first one focuses 
the discrepancy between users' intentions and actions under the influence of a given algorithm and without it. Put into the classical framework of the reasoned action approach \citep{fishbein2011predicting}, algorithms would be prone to modify the environment of users and possibly their skills: their intervention, if any, occurs between intentions and actions.
There appears to be at least two main ways of discussing behavioral divergences induced by algorithms and their magnitude. 
One consists in proposing, indeed, an extensive debate on the normative features which algorithms should conform to: for example, discussing whether such and such algorithmic principle or device guarantees ``equal'' access to such and such type of content. This question is strongly connected to the recently emerging and variously discussed notions of algorithmic ``fairness'', ``neutrality'', ``justice'', or ``partiality'', inter alia. Several scholars previously emphasized that the task of proposing a robust, operational definition for these notions could turn out to be rather difficult, if not impossible without providing further context and normative criteria \citep[see e.g.,][]{barocas2013governing}. This may indeed lead to a recursive debate, as it further requires to define the principles according to which ``equality'' is desirable (and which type thereof), or which bias would be undesirable (and along which dimensions). This, in turn, seems to necessarily involve external arguments. A second approach consists in appraising how algorithms may modify the inputs users are subjected to and, hence, their possible actions or beliefs. This does not necessarily mean reverse-engineering them or understanding how they were designed \citep{kitchin2017thinking}, but it seems to at least require to be able to describe a connection between input and output patterns, and to further differentiate this connection with regard to the amount of exposure to such and such algorithm (or to none, when this makes sense).
In other words, the point is to estimate the \emph{relative} impacts of such and such algorithm, with respect to one another as well as with respect to an organic, algorithm-free behavior of reference --- and thus to describe how it \emph{shifts} users away from that reference, and to which extent.\footnote{In this respect, the discussion in this paper will much less focus on specific algorithmic techniques (e.g., such and such type of machine learning) than on the stylized resulting effect that such and such algorithm induces (e.g., some form of reinforcement, some favoring of such and such type of content, etc.): first, there are various ways of implementing the same computational principle and it is to any extent difficult to know which precise technique has been actually employed by such and such platform, second, much more important is the possibility of distinguishing what happens in presence of such and such algorithm \hbox{vs.} in its absence.}  This leads us to distinguish situations where algorithms rather ``\emph{read}'' users' minds (small discrepancy) from those where they rather ``\emph{change}'' them (high discrepancy).

The second dichotomy is more instrumental and makes a distinction between two levels where algorithms may distort the stimuli that individuals are exposed to. The first level is  related to an information processing issue, essentially in terms of information reduction or projection. In effect, algorithmic principles encode a way of transforming large amounts of information into reduced subsets of cues, hints, pieces of knowledge, pointers, metrics, etc.  Put differently, this question focuses on the \emph{upstream} impact of algorithms in the \emph{rearrangement} of information from large databases before providing it to users.
The second level principally relates to psychological effects, \hbox{i.e.} how algorithms may affect the way individuals process the information they are provided with.  This question tends to focus more on the \emph{downstream} impact of algorithms in the \emph{arrangement} of information in its presentation to users, and leans more toward ergonomics and human-computer interaction issues. Both levels are admittedly intertwined. Their joint appraisal further relies on an interdisciplinary endeavor revolving mainly around contributions from computational and cognitive sciences. Besides, in both cases, we shall nevertheless show that social features play a key role, and lead us to emphasize the role of social cognition in algorithmic distortions --- either social cognition in the distributed sense, as the aggregation of behaviors of the underlying social system in which users evolve (e.g., a platform, the whole web), or social cognition in a more individual sense, \hbox{i.e.} how algorithms mobilize and plausibly influence some of the social cues which actors typically make sense of. 

The paper will be organized in three main sections: section~\ref{sec:romcom} will principally discuss the overarching dichotomy between reading or changing users' mind by illustrating its manifestations in the classical case of search engines. Section~\ref{sec:rearrangement} will then focus on the distortions induced by the algorithmic rearrangement of information, emphasizing specifically its locality and egocentricity in the case of affinity-based platforms (such as online social networks), while section~\ref{sec:arrangement} will deal with the potential effect on user behavior of the algorithmically-mediated arrangement of information.


\section{Read our mind \hbox{vs.} change our mind}\label{sec:romcom}

As suggested in the above, algorithmic influence on cognitive processes may first be appraised through a dichotomy distinguishing algorithmic devices which rather \emph{read our mind} (ROM) from those which tend to \emph{change our mind} (COM). More precisely, one may wonder whether algorithms merely identify our preferences and second-guess us, thereby acting as specialized cognitive helpers who essentially facilitate the otherwise taxing exploration of large amounts of information
; or if they actually behave as automatic gatekeepers which control our epistemic landscape and eventually guide and modify our behavior.  In turn, the ``COM'' cognitive bias should also be appraised with respect to how humans would already behave and proceed, without algorithms. In thus defining ``ROM'' as a baseline organic behavior under the smallest possible algorithmic constraints (namely: either no algorithmic information processing at all or a uniformly random one), we nonetheless disregard temporal effects: i.e., the same action occurring much earlier under algorithmic intervention than without it would still very much correspond to ROM.
In other words, the issue at stake in the ROM/COM dichotomy is whether algorithms speed up and expand the conditions of application of our free will or they rather limit or distort it.
The consequences of the latter appear to be of higher interest than the implications of the former, 
which may simply be about increasing information processing efficiency and productivity. 

\paragraph{A toy example: the case of search engines.}
We may start by considering the simple yet wide-ranging case of search engines. These devices are indeed one of the main entry points to online information which, for an increasing number of people, represents a significant share of the information that they access at all. 
Let us focus on the example of Google Search which provides an ubiquitous prototype of the issues raised by such engines and an excellent illustration of the two polar sides of the ROM~/~COM dichotomy.  Specifically, search engines feature two main classes of algorithms in order to guide users in their information queries: the first one consists of the semi-automatic completion of query words through suggestions based on what users are typing or typed.  The second one plainly corresponds to content ranking methods which are used to display the search results themselves.

\paragraph{Query expansion.} The first class is usually denoted as ``algorithmic search optimization'' and typically aims at expanding queries by proposing likely terms based on query logs made by other users \citep{baeza2005web}.  It induces two subcases: on the one hand, instantaneous term suggestion while typing, on the other hand, query correction after entering a first query. At face value, such mechanisms appear to work as a cognitive help in adaptively guessing what the user actually wants --- even more so when users carry out ``navigational queries'' (i.e., when they target a specific website or resource) rather than ``informational'' ones (when they look for some type of information without having a priori assumptions on which resources they should actually target) \citep{broder2002taxonomy}.

For one, instantaneous suggestion seems to purely target a productivity increase -- helping toward producing a list of query terms more quickly. For instance, an algorithm proposing ``Obama'' after a user started typing ``Barack'' would have fulfilled its role if Barack Obama was indeed the initially intended goal. automatic suggestion. In terms of the shape of the information landscape offered to users, such suggestions would have a neutral effect: users perform the same action with and without them, which appear as a purely ROM device --- they mainly affect the speed of human-computer interaction.
A string of debates nonetheless emerged regarding their possible diversion effect. 
\cite{baker-2013-why-do-white} specifically evoke the possibility of reinforcing discriminatory stereotypes when query completion suggestions correspond to classist, sexist or racist constructs. For instance they conjecture that, when users start to type ``white people are'', suggestions might reflect the most common queries and possibly prejudices, which are in turn displayed to all users.
However, it is unclear whether this phenomenon does actually distract a sizable portion of users from their initial search. A few empirical studies rather point to the opposite: \cite{shah-2013-exploration} demonstrate with a simple experimental protocol that the search process is not significantly affected, be it in terms of actual queries, visited web pages or sites, and transitioning from a query to another. \cite{mitra-2014-on-user} further show that users essentially rely on auto-completion when they have already typed half of their query characters and when they target rare terms, suggesting that this feature is mostly used to accelerate spelling rather than trigger unintended searches. Both studies tend to frame query expansion techniques as ROM algorithms only marginally diverting users.

Notwithstanding, the continuum between the two ROM and COM poles manifests itself in a more nuanced manner when we further consider the case of so-called ``related queries'' \citep{baeza-yates-2004-query}. The underlying principle is quite straightforward and relies on the assumption that distinct users who are actually interested in the same information may still formulate their query differently, either in terms of semantics (\hbox{i.e.}, formulating distinct phrasings for a similar request) or of focus (more or less complete phrasing for a similar request). To produce related queries, a typical approach consists in using the history of all queries which led users to click on a specific URL, which reveals a valid mapping from these queries to that digital resource \citep{tan-2007-term,zhang-2006-mining}. Queries pointing to the same URLs are thus considered equivalent and then clustered in order to propose either alternate phrasings or more precise queries, \hbox{i.e.} a more relevant term set in response to an initial query. As such, the goal of a query correction device appears again to consist in reading our mind better --- Google Search is almost explicit about this: ``Did you mean *** ?'', the engine asks candidly. Beyond the weak diversion potential of auto-completion, these techniques may still induce the additional side effect of channeling similar yet less conventional queries onto identical resources.
As \cite{baker-2013-why-do-white} rightfully point out, a query suggestion that a user perceives as relevant to their information quest would be increasingly considered to be adequate to the original query (irrespective of its distance or similarity to that query). 
In this respect, these techniques are likely to boost phrasings and/or results which previously suited a majority of users and to shift alternative queries toward more frequent ones, inducing a sort of one-result-fits-all-queries effect. In any event, within a certain domain, mainstream queries may be further reinforced and search diversity may be affected, which illustrates a possible drift toward the COM pole. 


\paragraph{Result ranking.}
The second class of search algorithms relates to the core of the engine itself: how to find, rate and sort results for a given query. The underlying principles induce a variety of cognitive biases which have also been discussed in a relatively abundant literature \citep[as early as][]{introna-2000-shaping}. Indexing the web is an endeavor as old as the web itself: it started manually through the development of so-called ``weblogs'', which were basically logs of web browsing sessions published by individual users \citep{siles-2011-online} (wherefrom the word ``blog'' originates) and, later, with dedicated platforms gathering categories of bookmarks (such as Tim Berners-Lee's Virtual Library, introduced concurrently with the invention of the web, \hbox{Yahoo!} or the collaborative Open Directory Project). The exponential growth of web content pressed the emergence of dedicated, automated content indexation and curation platforms. These algorithmic gatekeepers have progressively supplemented and partly supplanted human gatekeepers in accessing digital information \citep{hargittai-2000-open}. 

Despite that, automation did not remove all of the subjectivity that could govern the earlier human curation efforts:  it also derives from man-made algorithmic principles whose effects remain to be appraised, even as of today. 
The earliest successful search engines (including Excite [1993], Webcrawler [1994], Lycos [1994], Infoseek [1995], Altavista [1995]) were rather based on semantic features, a principle inherited from classical database management criteria: information indexation and retrieval essentially relied on lexicometric concepts, such as the frequency of term use in a web page or in its meta-data. Documents whose lexical features match query terms pretty well are considered relevant to the query \citep{gloggler2013onthepage}. 
By contrast, Google [1998] introduced through the famous ``\hbox{PageRank}'' algorithm a notion of relevancy based on structural features of the hyperlink topology of the Web and, more interestingly and more broadly, to properties which are external to the requested documents: a resource is deemed pertinent for a given query if many webpages cite that resource all the while mentioning the requested query terms --- information is thus filtered and ranked according to a form of distributed vote produced by the Web, construed as a huge and implicitly crowdsourced rating system.  The canonical (and perhaps ideal) Google Search algorithm does (or did) embed and implement a social cognition principle: relevance is based on the judgment of other web content producers, thereby favoring content enjoying a socially distributed support rather than exhibiting a semantically valid match.  The analogy is also that of a randomly browsing user who would more often than not end up on webpages featuring a higher number of incoming links, in a recursive fashion \citep{brin1998anatomy}; the algorithm would thus give these webpages a higher rank.  Whichever the interpretation, the risk of reinforcement dynamics is plausibly more present than in the semantic case, for users and web content producers alike will be preferably channeled to well-connected webpages: \cite{hind-goog}, for one, were among the first ones to underline this type of filtering which they denoted as ``googlearchy''. It could reinforce preexisting dominant positions by giving more visibility and distorting further the subsequent epistemic landscape toward sources or content which are already heavily cited, hence visible. Moreover, very much like in the query expansion situation, valid matches between a query and a URL, evidenced by user clicks, likely contribute further to a higher ranking of that URL in reply to the query.

\bigskip
In some way, facilitating the exploration of a huge database ---the web--- by returning documents which match some logical criteria ---presence of such and such term--- is akin to a 
cognition-extending device geared at easing
the discovery of a needle in a digital haystack. It thus seem to lean toward the ROM pole, markedly so when results are simply sorted with respect to term frequency. Here, a large database query system principally carrying out term matching would correspond to a case of minimal algorithmic intervention and would arguably represent a good ROM reference. By contrast, more sophisticated computations as achieved by Google's \hbox{PageRank} induce a non-trivial re-ordering of information which is likely to favor some type of content in a dynamic and reflexive way. Its impact thus remains pretty difficult to appraise, all the more so from a user perspective.

More broadly, it is clear that a number of information ordering and filtering principles may shift algorithms toward the COM pole, combining semantic, stuctural and temporal features in a way that is generally not publicized and whose effects are consequently often hard to assess. 
The influence of algorithmic devices in this regard may be studied along two key dimensions, usually in sequence: on the one hand, the \emph{rearrangement} of information, \hbox{i.e.} how information is first being selected and filtered by algorithms (section~\ref{sec:rearrangement}), and then, on the other hand, its \emph{arrangement}, \hbox{i.e.} how it is conveyed to users (section~\ref{sec:arrangement}).

In the simple case of search engines, again, the first dimension may further be illustrated by the fact that even the same query may lead to a quite significant diversity of results, thus revealing a non-trivial process of information selection that might go far beyond the above-discussed issues. Using two of the most used search engines, Google and Bing, \cite{dillahunt-2015-detecting} experimentally demonstrated that an identical request made from distinct browsers, all in private session mode (\hbox{i.e.} with as little personalization as possible) already yields distinct subsets of resources. The second dimension can be exemplified by the guidance entailed by the way results are displayed, especially the so-called position bias \citep{dupret2008user}, which diversely corresponds to the decaying probability of examining results down the list \citep{Joachims:2007:EAI:1229179.1229181} or the increased likelihood of clicking on a document next to a very relevant one.

We shall examine in detail what is currently known about the cognitive biases induced by the algorithmic rearrangement and arrangement of information respectively in section~\ref{sec:rearrangement} and in section~\ref{sec:arrangement}.



\section{Ego-centered rearrangement of information: affinity-based platforms\label{sec:rearrangement}}

The combination of principles into distinct algorithmic recipes each lead to specific informational landscapes: the Web, for one, may obviously be rendered differently by distinct platforms or ranking engines; content relevance varies accordingly.
This is all the more pregnant in the case of so-called affinity-based platforms where the construction of these landscapes are further altered by user behavior --- more precisely, where algorithms learn from user preferences to feed them with content based on their prior preferences. In this case, the reinforcement dynamics evoked in the previous section do not only appear at a rather global level, in a manner roughly similar for all users, but they are also significantly customized for each user. At the user level, this kind of ego-centered positive feedback loops may contribute to enclose users in what has been first denoted as ``filter bubbles'' by \cite{pari-filt} and has been described later as ``selective exposure'' \citep{baks-expo}. At the system level, the interplay between customization and algorithms makes it generally harder to evaluate the direct effect or contribution of some algorithmic principles, since this requires not only a good analytical understanding of such principles but also a global empirical knowledge of the configuration and distribution of user preferences. What is at stake is the algorithmic interplay between the local, ego-centered preferences and actions, and their global aggregation, which feeds back into the individual level.

\smallskip
Preferences can be \emph{explicit} when users willingly declare their affinity for certain type of content, for instance by subscribing to the publications of other users or indicating interest for specific groups, collectives, cultural artifacts (as is the case on Twitter or Facebook), topics or keywords (for instance for the provision of taylored news on DailyMe). Besides, previous user behavior and choices are also tracked by platforms in order to build up individualized profiles of \emph{implicit} preferences, as is for example the case on YouTube with videos which are labeled as ``recommended for you''.
Both implicit and explicit data are used to shape automated recommendations, often with the help of machine learning techniques. 

Obviously, the intended objective leans again toward ROM: from a functional viewpoint, the ideal situation would be to have just enough personalized data to automatically provide users with the items they would have been the most eager to consume had they had enough time to scout the available content. This is most visible when platform algorithms trivially respect preferences indicated by users, as is the case for instance with so-called news feeds based on Twitter follower networks or Facebook friendship networks: in this case, information is principally selected and filtered from a preselection of sources voluntarily provided by users, whereby the platform simply limits the informational horizon to these sources with the tacit agreement of users. 

\paragraph{``Human'' \hbox{vs.} man-made algorithms.} In other words, such platforms principally appear to assist the implementation of the pre-selection process desired by human actors. In this regard, what is being described as algorithmic filter bubbles could hence be understood as little more than automated helpers bringing users more quickly and more likely to the content they were looking for anyway --- once again a cognitive accelerator for navigating more easily the socio-cognitive bubble that users would have otherwise naturally evolved in, only more slowly and less efficiently. For instance, if we consider the emergence and permanence of cohesive clusters of links and information in online ecosystems, we may first have to attribute their existence, and the induced narrowing of the cognitive landscape, to \emph{``human'' algorithms} rather than \emph{man-made algorithms}, which thus reflect (and mechanically enforce) the organic limited rationality that actors may want to impose on themselves, rather than create it. Put differently, the organic selection \& influence process long identified in the sociological literature, whereby peer influence coevolves with peer selection and jointly reinforces it, applies way before any algorithmic device is put in place.  In effect, many online social systems spontaneously exhibit a fragmented and clustered structure (observed early on blogs \citep{adam-poli} and similarly in micro-blogs  \citep{liet-when,barbera-2015-twe}), where interactions seem to be heavily influenced by homophily \citep{cono-poli} and selection \& influence processes \citep{lewis-2012-social}, absent algorithms.

Notwithstanding, the possibility that such algorithms tend toward COM resides in the manner they could still interfere with this organic behavior. 
At first sight, the magnitude of COM has to be evaluated in terms of what user intentions and actions were without resorting to the algorithm \hbox{vs.} what they are with it. Admittedly, the intended goal of recommendation on online platforms generally consists in maximizing user participation and satisfaction in the broad sense. Yet, it is also unclear whether this optimization is aligned with prior user interests. Consider the case of YouTube again: its recommendation algorithm aims at maximizing view time \citep{covington2016deep}, but the extent to which this goal corresponds to what users may have initially been willing to consume on the platform remains relatively unknown.  Being hooked on a series of videos recommended by the platform following, say, a search query, comes to adopting a form of algorithmic serendipity which may have little to do with the initial query or even with the organic serendipity subsequent to that query.

\paragraph{Algorithms and serendipity.}
More than anything, one of the key current empirical questions is whether algorithmic recommandation leads to a narrower and less diverse information and interaction space than would otherwise be the case --- \hbox{i.e.} appraising whether the information, books, movies, songs we consume would have been more diverse without algorithmic mediation. A stream of studies has indeed been concerned over the last decade on the effect of recommandation algorithms on information consumption diversity. The commonsensical view is that algorithms mainly strengthen the organic bubble that users are inclined to create. 
However, on the whole and maybe counter-intuitively, this literature seems to demonstrate that algorithms tend to expand rather than limit the cognitive horizon of users. This phenomemon has been variously observed on video sharing and streaming platforms, movie or blog recommender systems, online social networks, to cite a few.

For instance, in their study of viewing behavior on YouTube, \cite{zhou2010impact} start by showing that videos that are referred to from videos with a higher number of views are likely to feature more views as well: recommendation seems to sustain a form of reinforcement dynamics --- rich recommends rich. Yet, and perhaps more importantly, they also demonstrate that the presence of recommendation tends to increase the heterogeneity of the consumption portfolio \emph{at the global level}, \hbox{i.e.} views are distributed more flatly and more equally across all videos: recommendation sustains a form of equalizing redistribution. To reconcile these two apparently conflicting observations 
we might hypothesize the following explanation. By default, we can suppose that users tend to land proportionally more on videos with higher numbers of views, obeying a Markovian process whose transition law follows the prior distribution of views. In turn, the fact that aggregate audience levels are more balanced among videos when recommendation is active may suggest that the algorithm has a propensity to somewhat redispatch viewers from popular videos toward less popular ones. This is consistent with the a posteriori observed correlation between referred and referring videos in terms of audience size. In a nutshell, in front of a mainstream-friendly behavior, recommendation appears to pull users toward more niche content, hence increasing informational diversity at the system level.

More recently, \cite{aiel-evol} focused on Tumblr, a blogging platform where posts mainly consist of photos and optionally feature a text caption. Users can subscribe to posts of other user accounts whom they discover (1) either themselves, plausibly by wandering from blog to blog, or (2) by following recommendations provided by the platform. The study shows that recommendations lead users to build up a more diverse portfolio of subscriptions. In fact, it appears that users organically tend to add new subscriptions by iteratively exploring the immediate neighborhood of their network, which is in line with earlier works on ego-centered network formation. \cite{libe-link} show for instance that ``organic'' links (\hbox{i.e.} link creation that is not mediated by recommender systems) tend to be best predicted by focusing on individuals that have a lot of neighbors in common, \hbox{i.e.} who belong to densely-knit user clusters surrounding ego. On the contrary, algorithmic suggestions point at highly connected users, in relative terms, who are generally located farther in the network than the local neighborhood of ego. In other words, while users seem to be rather niche-friendly in topological terms (\hbox{i.e.} they explore their own local social neighborhood), by contrast with the previous case recommendation directs them toward more mainstream sources, hence increasing informational diversity as well, although through different means.

Similar conclusions with respect to the interplay between serendipity and reinforcement dynamics may be drawn from further studies on cultural recommendation. \cite{nguyen2014exploring} examine a platform which implements collaborative filtering (called MovieLens), \hbox{i.e.} a recommendation system based on the assumption that a user previously interested in the same items as another user may be interested in some of the other items of interest to that user. What sounds like a reinforcement principle actually leads to an increase of genre diversity in the list of watched movies for users who adopt these recommendations with respect to those who do not.
In a different context, \cite{datta-2018-changing} focus on data describing traces of music consumption in a cross-platform setting. More precisely, they check whether Spotify, a music streaming service which partly features algorithmic recommendation (such as the automatic continuation of playlists or of listening sessions) enables users to consume more diverse and newer music than those using platforms based on catalog search only (\hbox{i.e.} where users may only look for specific content, such as iTunes). They observe a significant, important and persistent increase in the number of unique artists, songs or genres that are listened to by adopters. The exact principles which drive Spotify's recommender systems are naturally not public and perhaps not even entirely understood. It nonetheless appears that this kind of recommendation usually relies in part on previous user behavior, and in part on similarity-based, frequent pattern mining and collaborative filtering algorithms \citep{bonnin-2015-automated}, \hbox{i.e.} it makes use of the typical associations between artists or songs that other users did.

\paragraph{Human with man-made algorithms.}
A recent study by \cite{baks-expo} is particularly interesting in that it both touches the ROM-COM divide and addresses the compound and intertwined effects of human behavior and algorithms. The authors look at the complex effect of Facebook's ``news feed'' algorithm on the diversity of the informational space displayed to users. In practice, the study illustrates how three sequential filtering operations influence information consumption, namely:
\begin{enumerate}[(i)]
\item the explicit preferences and information space limitation induced by the prior selection of sources by users (in terms of other users [or ``friends''] and of user groups / webpages of interest on the platform);
\item the algorithmic selection that the platform applies on all the information published by these sources; 
\item and, eventually, the actual instantaneous consumption of information, which is done by users, again.
\end{enumerate}
Both (i) and (iii) relate to organic behavior: the ego-centered network of Facebook friends is generally populated by offline acquaintances \citep{dunbar-2015-the-structure} and thus reflects a social selection process generally devoid of algorithms, while the action of clicking on some information on the news feed is also naturally triggered by users. Only (ii) may be ascribed to the selection induced by the news feed algorithm. \cite{baks-expo} essentially check the cumulative role of each step in exposing users to cross-partisan information, \hbox{i.e.} information which has a political valence opposite to that of the user. Given this process, they show that the selection algorithm itself has only a very moderate effect in the reduction of cross-partisan information available to users, with respect to what is already available in their (organically biased) ego-centered network. In more detail, they demonstrate that, say, conservative users generally have more conservative friends (their organically-built landscape of sources is biased) so that less cross-partisan information will have a chance to reach them, already before the algorithm applies. For a given news feed, \hbox{i.e.} after the algorithm applies, they also do select more conservative information (their instantenous consumption behavior is biased) --- both well-known organic biases.  Hence, the reduction of diversity appears to be principally due to human pre- and post-selection, which does not support the existence of a change-our-mind effect. Rather, it is possible to argue that the algorithm might be very close to the read-our-mind pole. Indeed, while it certainly learns from user previous behavior and thus seems to be designed to reinforce it, it may just \emph{help} users to sort out and filter what they are already most inclined to consume among the huge amount of information published by their ``friends'', while it does not change much what they would have eventually consumed if they were confronted to all their publications in an unfiltered fashion (at least in terms of cross-partisan information).\footnote{To bring a definite answer to the ROM-COM distinction in this case, it would have been fruitful to introduce a null empirical protocol where the algorithm would be absent and then compare the eventual reduction of diversity of (i)+(iii) with that of (i)+(ii)+(iii). Indeed, while (ii) does not seem to reduce the amount of partisan \hbox{vs.} cross-partisan information from (i), it may still select partisan information that is most effective when it reaches the user, so that information selected in (iii) is even more partisan that without (ii), all other things being equal.}

\smallskip 
In all these cases, we witness the possibly paradoxical situation where algorithms rely on various acceptations of some form of reinforcement yet actually tend to open up the available horizon and thus support the exploration of novelty. The key source of diversity here lies perhaps in the fact that recommendation is partly based on the aggregation of individual behavior, \hbox{i.e.} that the reinforcement operates on signals available at the collective level --- which seems to be eventually beneficial to individual serendipity.  All in all, it appears that users rather than algorithms are one of the main sources of limitation of the cognitive horizon when information rearrangement is based on user preferences.

\paragraph{Algorithms masquerading as humans.} Let us finally evoke a perhaps more sophisticated form of socially-mediated algorithmic rearrangement of information. It is due to an entirely different kind of algorithms: robots, also simply called ``bots'', whose role is to imitate human agency. By contrast to algorithms explicitly devoted to filtering and recommanding information, these devices introduce non-human actors \emph{on par} with human actors on the platforms --- in that they are theoretically and technically capable of the same actions: relaying information, expressing opinions, rating content and other actors. As such, they contribute to modifying user behavior in a somewhat external fashion with respect to the algorithms directly running on the platform they populate : bots create information and establish interactions rather than ``just'' filter them.

While the contribution of online bots may sometimes be construed as a positive help toward social cognition processes \citep[for example algorithmic governance in Wikipedia, see][]{mull-work,nied-wisd,Steiner-014}, the emphasis is commonly put on their possible role in distorting digital spaces, with sometimes concrete real-world effects \hbox{e.g.}, tampering with social unrest, stock markets or elections \citep{thomas-2012-adapting,ferr-rise}. In this latter case, bots may principally exert two different types of actions. 
First, by interacting indirectly with other algorithmic devices \hbox{e.g.}, by altering the reflexive counters of content appreciation, such as so-called ``trending topics''. The global information landscape may for instance be modified by flooding a platform with irrelevant or malicious information, through coordinated attacks aimed at passively exerting censorship \citep{thomas-2012-adapting}.
Second, by interacting directly with other users, thus directly modifying part of their own information landscape. 
Nonetheless, even if bot activity may be significant on a given platform \citep{chu2010tweeting,lee2011seven}, their impact on a user's informational portfolio could remain much weaker than that produced by other human actors if bots are weakly entangled in that user's social fabric: for example, humans form the vast majority of Facebook friend networks. In other words, if bots talk to bots, their influence is moot. Yet, not only can the opposite be said on some platforms such as Twitter \citep{ferr-rise} but, as \cite{shao2018spread} show, social bots may appear to have a strong influence on content spreading. They demonstrate in particular that, while human users are responsible of the largest proportion of retweets, they still relay information from a significant proportion of bots.  Furthermore, bots need not be very sophisticated to become connected with human users, as shown by \cite{bosh-soci} on Facebook or \cite{aiel-peop} on aNobii (a social network platform centered around book dicussions) --- their online experiments involved basic social phishing behavior respectively through random ``friendship'' requests or slightly more elaborate machine-learning-based prodding behavior. In both cases, (seemingly) human users responded positively to these solicitations with a surprisingly high rate (generally more often than not). Of course, the recent advent of sophisticated bots based on advanced machine learning and natural language processing techniques is likely to improve their agency and thus manipulation potential. The much-discussed case of Microsoft's twitter chatbot ``Tay'', which elicited a massive number of conversations (about 93k tweets in about 16 hours), illustrates this issue perfectly. Even if bots are based on apparently innocuous principles --- here, maximizing ``human engagement'', \hbox{i.e.} triggering as many interactions with other users as possible --- the potential nocivity of their influence remains to be appraised also in regard to how humans tend to assign human agency to these bots \citep{neff-2016-talking}. This brings us to discuss more broadly the other side of the cognitive effects of algorithms, namely the shape of human reactions to algorithms and algorithmic guidance.



\section{Information arrangement and reactions to algorithmic guidance\label{sec:arrangement}}

We now focus on the rather ``downstream'' side of algorithmic guidance. Several experimental protocols have been devised to study how algorithmic choices in the restitution of information affect relevance judgments, clearly assuming that front end aspects contribute a priori more to COM than ROM.\footnote{To discuss ROM \hbox{vs.} COM in this regard, we may for instance simply think of a plain list of items ordered in a uniformly random fashion.} Turning again to search engines, \cite{epst-sear} demonstrated that political opinion may be manipulated by the simple reordering of information returned by a given query. In their experiment, US participants were first asked to read a short biography about two main candidates in an Australian national election and provide their a priori impression of them. They were then put in front of a mock search engine displaying the same results for all participants, except that the presentation would vary according to three experimental conditions: the display of results would be ordered in favor of one candidate, the other one, or none in particular. After freely navigating the results for up to fifteen minutes, participants were asked again about their opinion. It is well-known that users tend to proportionally pay much more attention to the links at the top of the page and this phenomenon was also reproduced in this experiment. The authors could notice a significant shift in preferences in a direction that was consistent with the reordering favoring a given candidate, irrespective of the initial opinion. Not only did the above-mentioned position bias modify the likelihood of being confronted with some information, but it had a significant effect on the formation of beliefs, especially in contexts where moderate changes in the affected population may entail an important political impact.

\paragraph{Promoting popular \hbox{vs.} diverse content.\remove{[REARRANGE WHOLE SUBSECTION]}}
Understanding the impact of the algorithmic arrangement of information makes it also possible to reverse engineer user behavior, especially their appetence toward diversity. For one, \cite{park-2009-newscube} examined the effect of grouping news items related to the same underlying event, \hbox{i.e.} recommanding similar articles together under a main entry rather than displaying them separately on the same level. They showed that the grouped display favors both the perception and the consumption of diverse viewpoints. In other words, in response to a query on some current event, showing that there were many possible options which seemed to increase curiosity.

A key related issue concerns the diversity in the popularity of content and, 
implicitly, user preference for mainstream \hbox{vs.} niche content. Are users going to react differently to popular content ? Let us first consider what we know about the computer-mediated access to content, without algorithms. We have to mention here the 
natural tendency of many social systems to exhibit a hierarchical structure where few receive a lot of attention and many receive little, which \cite{shirky2006power} links in part to the existence of free choice:
\begin{quote}
\small\em In systems where many people are free to choose between many options, a small subset of the whole will get a disproportionate amount of traffic (or attention, or income), even if no members of the system actively work towards such an outcome. (...) The very act of choosing, spread widely enough and freely enough, creates a power law distribution.
\end{quote}
In a context of free choice and given this a priori heterogeneous distribution of attention, the Internet has nonetheless often been framed as a diversity-enhancing technology, in particular by making available content which benefits from little to no attention. This is also in part one of the core design features of the Internet: precursors such as \cite{lick-comp} envisioned communication devices as enablers of an easier access to specialized content and interlocutors that would normally be hard to find.
\cite{anderson-2004-the-long} hypothesized that the online accessiblity of otherwise unreachable items would trigger an increase in the consumption of niche content and thus, system-wide, an increase of diversity. This intuition has been partly confirmed by data on the evolution of the distribution of attention for high- and low-selling videos and DVDs \citep{elberse-2006-superstars} and later on the online consumption of various types of cultural items (movies, music and web sites \citep{goel-2010-anatomy}): niche content benefits from the larger and wider availability provided by online channels. Yet, at the same time, it appears that consumption also increases for the most mainstream content: the top of the distribution becomes more concentrated, i.e., fewer highly popular titles make the bulk of sales than before. \cite{elberse-2008-should} specifically distinguishes light users who happen to be mostly focusing on mainstream content from heavy users who tap into both ends of the distribution, including the long tail.

Measured user attitudes toward algorithmic recommendation seem to reflect these ambiguous results. 
\cite{steck-2011-item} observes that many recommender systems traditionally induce a negative bias toward the least popular items, first as a result of the fact that available data on items follows a power-law pattern too and is thus typically scarce in the long tail as well. The author proposes a method to correct this preference for mainstream content by promoting a variable amount of items stemming from the long tail, thus biasing the displayed results in favor of less popular items. He follows up with an experimental study testing the biased against the non-biased algorithms. It shows that users tend to slightly prefer the former, while they also seem to need familiar (popular) recommendations from the mainstream in order to be satisfied with the algorithm results. In other words, users seem to appreciate surprising or less known results inasmuch as they are collocated with more expectable ones --- \hbox{i.e.} displaying content from both ends of the distribution.

\paragraph{Adding reflexive signals.}
Popularity may both be intrinsic (it directly corresponds to the actual size of the interested audience) and perceived (it may correspond to user beliefs or expectations about the interested audience). Here too, there exist specialized algorithmic devices whose aim is to convey a reflexive information about the system-wide popularity of some items, thus influencing the perception of users of their relevance.
The pioneering work of \cite{salg-expe} demonstrated how peer influence mediated by algorithmic rankings may modify the judgments on cultural quality.  Their experimental setting was based on an artificial musical market where participants could browse and rate (like or dislike) previously unknown songs in three distinct experimental situations featuring an increasing level of social influence. In the first (control) situation, participants could only see band names and song titles. In the second one, they were shown the real-time popularity of songs (number of downloads) within the experiment. In the third one, songs were ranked and sorted according to popularity. Popularity was shown to be only partly influenced by intrinsic quality, in the sense that the very best songs were rarely unpopular while the worst ones were rarely very successful. However, in both social influence conditions, final song popularity was shown to be generally unpredictable, \hbox{i.e.} it seemed to depend only weakly, if at all, on intrinsic quality. What is more, a corollary of this phenomena is that final rankings were not stable across social influence experiments: they appeared to be depend essentially on each realization of the experiment and its own intrinsic reinforcing social dynamics, very much like a generalized Polya urn process where any outcome is equally likely. 

Similarly, \cite{messing-2014-selective} focus on the reflexive effect of displaying social endorsements close to a given information. They devise an experimental protocol where pieces of news are displayed along with two informations that the authors may manipulate: the source name and a number indicating how many users recommended that specific content on a social media platform (Facebook). Sources are identified as either left-wing or right-wing leaning (e.g., MSNBC \hbox{vs.} Fox News). The orientation of participants is also known a priori. This enables them to contrast the combined effects of homophily (left-wing users may click preferentially on news items displayed as originating from left-leaning sources) and social endorsement (users may prefer items having received many recommendations). They show that the display of social endorsement predicts much more effectively selection by users and even dwarfs the effects of partisan selectivity. 

In both cases, displaying a reflexive computation on the social system appears to channel collective opinions towards a relatively random attractor which may have little to do with independent quality judgments. Put differently, these results relate to a form of passive algorithmic nudging of users through the reflexive display of information about the whole system --- they more broadly show that information and interface design may influence belief formation, especially as this kind of reflexive algorithmic device tend to be widespread (think of online media websites displaying in real-time the most popular or trending articles).  Going further, some scholars advocate the manipulation of information arrangement to fulfill certain normative goals, preferably in a virtuous manner. In this regard, \cite{resnick-2013-bursting} review various algorithmic tricks aimed at increasing exposure to diverse information. This includes in particular the possibility of temporarily displaying information the way it is seen by other (and ideally different-minded) users on a given online platform\footnote{Various helper tools have recently been developed to concretely implement these ideas, such as FlipFeed, which computes and displays Twitter feeds from likely politically opposed users, or    PolitEcho, which shows Facebook users the distribution of the likely political orientation of their friends.}
as well as, for instance, displaying viewpoints expressed around a topic of interest by other users on the whole web \citep{murakami-2010-statement}.

\paragraph{Variety of user attitudes toward algorithms.}
Most of these studies implicitly assume the existence of an average user, \hbox{i.e.} results are discussed on the basis of an average behavior and attitude toward the effect of algorithms.  Some authors endeavored at distinguishing several subpopulations of users who are more or less sensitive to algorithmic guidance, or react differently.  This may be somewhat basic: for instance, \cite{chen-2011-speak} demonstrate that people have distinct uses of a same device, distinguishing Twitter users who seek social interactions from those who seek information, and demonstrating that algorithm performance differs (statistically) significantly between the two groups. \cite{munson-2010-presenting} design an experiment similar to several above-mentioned studies, where participants are shown news item corresponding to a political orientation aligned with theirs or not, in an algorithmic curation context resembling the display of a typical search engine. They find behaviors which are consistent with those of \cite{park-2009-newscube} in that some users may prefer to see diverse collections of items. However, they manage to differentiate challenge-averse and diversity-seeking users, demonstrating that challenge-averse individuals would prefer seeing agreeable items only, a feature which may put off diversity-seeking people. In turn, this suggests designing distinct ways of presenting results depending on the audience, \hbox{i.e.} tayloring the interface and possibly the algorithm to the alleged or infered type of users --- a form of meta-personalization. Building upon this, \cite{munson-2013-encouraging} further devise a web browser extension in order to nudge purported challenge-averse users (or rather, in this context, users overwhelmingly consuming news of a given political orientation). The algorithm displays in real-time the imbalance of a participant's news diet, with the normative expectation that users would be inclined to correct their consumption behavior towards a more politically balanced portfolio of articles. They show that only some users do indeed equilibrate a bit more their habits, while many are not affected, which may be interpreted in a consistent manner with the above observations on the existence of two challenge-averse and diversity-seeking subpopulations.

Further, at a more meta level, users may also be aware of or hypothesize differently the functioning of the algorithmic guidance they are subjected to, forming different beliefs in this respect --- what I would denote as \emph{folk algorithmics}, by analogy with folk psychology, biology or physics. 
This last strand of research corresponds to a barely emerging literature. Recently, \cite{rader-2015-understanding} conducted a query-based study on about 500 Facebook users to appraise their perception of the  algorithmic curation exerted by the News Feed device. Participants were asked to answer a simple question, ``does FB show you all ?'' [Yes/No/Maybe] and subsequently provide a paragraph-based justification.
They extract an (overlapping) typology of four user attitudes: ``passive consumers'', \hbox{i.e.} users who have no clue (which concerns about 20\% of users), ``manual control'', \hbox{i.e.} users who assume that their friends control the audience of their publications when they don't see their posts (\char`\~25\%), ``symptoms of curation'', \hbox{i.e.} users who recognize some signs of algorithmic curation (especially because they unexpectedly notice the absence of some posts) (\char`\~80\%), and ``speculating about the algorithm'', \hbox{i.e.} users who attempt to form intuitions about the principles of the algorithms (\char`\~45\%). In the near future, the discussion on algorithmic distortions would likely greatly benefit from such informed typologies on user attitude towards algorithms, most notably by differentiating results with respect to the level of reflexive knowledge and mastering by users of the algorithmic guidance they are subjected and receptive to.

\section*{Concluding remarks}

This paper attempted to provide an overview of the influence of algorithms on cognitive processes by focusing on their deployment on online platforms and tools, impacting a significant part of users' informational and interactional landscapes.  I proposed to analyze these phenomena by developing a double dichotomy. The first one, so-called ``read-our-mind'' \hbox{vs.} ``change-our-mind'', addresses the discrepancy between users' intentions and expected actions under the influence of some algorithms and without them. The second one addresses the algorithmic influence on the prior information rearrangement \hbox{vs.} the posterior information arrangement. This made it possible to emphasize the need to identify the organic, baseline behavior of users as a key contribution to this discussion. Focusing on the issue of the potential limitation by algorithms of the cognitive horizon available to users, we could exhibit their potentially counter-intuitive role of supporting rather than hindering serendipity.  

This relies, in part yet crucially, on the fact that algorithms carry out a form of social aggregation of behavioral traces which contributes to relevantly expand the informational perimeter that users would normally access, absent algorithmic mediation. On the whole, the important conclusion that may be drawn from much of the current state of the art could be that algorithmic cognitive biases cannot be appraised without taking into account the underlying social processes which algorithms are building upon.
The point here more precisely relates to the algorithmic reification of existing social processes --- be it the existence of stereotypes (and, thus, their algorithmic reinforcement or stabilization) or the selection-and-influence chicken-and-egg sociological issue (and, thus, their algorithmic automation and perhaps invisibilization).

Further, what may also be needed in this debate is a change of perspective on algorithmic recommendation and specifically what users generally expect from this kind of devices.  More precisely, that users rely on algorithms not so much for reading their mind  than for asking them to change their mind: beyond the case of pure database navigation tasks (which search engines exemplify very well), users are best satisfied by an algorithmic bookseller rather librarian, giving them surprising (yet relevant) advices rather than reminding them about or guiding them toward familiar things.

The principles behind this bookseller are hard both to identify and to define. A possibly fruitful proposition in this respect could consist in researching the prospects and effects of opening algorithms in a both interactive and stylized fashion \citep{ekstrand-2015-letting}, \hbox{i.e.} providing them with some leeway to users to understand the general principles of the algorithms in use, the main \emph{differential} effects that they may induce on their baseline behavior, and perhaps most importantly be able to make choices as to the activation and weighting of the various ingredients that form a given algorithmic recipes.  Users would thus expand or constrain their own informational landscape in a more conscious manner, thereby given them back some free will while using algorithms.

\subsection*{Acknowledgements}
This paper has been partially realized in the framework of the ``Algodiv'' grant (ANR-15-CE38-0001) funded by the ANR (French National Agency of Research).

\footnotesize


\begin{thebibliography}{}

\bibitem[\protect\astroncite{Adamic and Glance}{2005}]{adam-poli}
Adamic, L.~A. and Glance, N. (2005).
\newblock The political blogosphere and the 2004 {U.S.} election: divided they
  blog.
\newblock In {\em LinkKDD '05: Proc. 3rd Intl. Workshop on Link discovery},
  pages 36--43, New York. ACM Press.

\bibitem[\protect\astroncite{Aiello and Barbieri}{2017}]{aiel-evol}
Aiello, L.~M. and Barbieri, N. (2017).
\newblock Evolution of ego-networks in social media with link recommendations.
\newblock In {\em Proc. 10th {ACM} Intl. Conf. on Web Search and Data Mining},
  WSDM '17, pages 111--120, New York, NY. ACM.

\bibitem[\protect\astroncite{Aiello et~al.}{2012}]{aiel-peop}
Aiello, L.~M., Deplano, M., Schifanella, R., and Ruffo, G. (2012).
\newblock People are strange when you're a stranger: Impact and influence of
  bots on social networks.
\newblock In {\em Proc. 6th ICWSM International AAAI Conference on Web and
  Social Media}, pages 10--17. AAAI press.

\bibitem[\protect\astroncite{Anderson}{2004}]{anderson-2004-the-long}
Anderson, C. (2004).
\newblock The long tail.
\newblock {\em Wired Magazine}, 12:170--177.

\bibitem[\protect\astroncite{Baeza-Yates}{2005}]{baeza2005web}
Baeza-Yates, R. (2005).
\newblock Web usage mining in search engines.
\newblock In {\em Web mining: applications and techniques}, pages 307--321. IGI
  Global.

\bibitem[\protect\astroncite{Baeza-Yates et~al.}{2004}]{baeza-yates-2004-query}
Baeza-Yates, R., Hurtado, C., and Mendoza, M. (2004).
\newblock Query recommendation using query logs in search engines.
\newblock In Lindner, W., Mesiti, M., T{\"u}rker, C., Tzitzikas, Y., and
  Vakali, A., editors, {\em Current Trends in Database Technology - EDBT 2004
  Workshops}, volume 3268 of {\em LNCS}. Springer.

\bibitem[\protect\astroncite{Baker and Potts}{2013}]{baker-2013-why-do-white}
Baker, P. and Potts, A. (2013).
\newblock 'why do white people have thin lips?' google and the perpetuation of
  stereotypes via auto-complete search forms.
\newblock {\em Critical Discourse Studies}, 10(2):187--204.

\bibitem[\protect\astroncite{Bakshy et~al.}{2015}]{baks-expo}
Bakshy, E., Messing, S., and Adamic, L.~A. (2015).
\newblock Exposure to ideologically diverse news and opinion on facebook.
\newblock {\em Science}, 348(6239):1130--1132.

\bibitem[\protect\astroncite{Barber{\'a} et~al.}{2015}]{barbera-2015-twe}
Barber{\'a}, P., Jost, J.~T., Nagler, J., Tucker, J.~A., and Bonneau, R.
  (2015).
\newblock Tweeting from left to right: Is online political communication more
  than an echo chamber?
\newblock {\em Psychological Science}, 26(10):1531--1542.

\bibitem[\protect\astroncite{Barocas et~al.}{2013}]{barocas2013governing}
Barocas, S., Hood, S., and Ziewitz, M. (2013).
\newblock Governing algorithms: A provocation piece.
\newblock Available at SSRN: https://ssrn.com/abstract=2245322 or
  http://dx.doi.org/10.2139/ssrn.2245322.

\bibitem[\protect\astroncite{Bonnin and Jannach}{2015}]{bonnin-2015-automated}
Bonnin, G. and Jannach, D. (2015).
\newblock Automated generation of music playlists: Survey and experiments.
\newblock {\em {ACM} Computing Surveys {(CSUR)}}, 47(2):26.

\bibitem[\protect\astroncite{Boshmaf et~al.}{2011}]{bosh-soci}
Boshmaf, Y., Muslukhov, I., Beznosov, K., and Ripeanu, M. (2011).
\newblock The socialbot network: when bots socialize for fame and money.
\newblock In {\em Proc. ACSAC '11 27th Annual Computer Security Applications
  Conference}, pages 93--102. {ACM}.

\bibitem[\protect\astroncite{Brin and Page}{1998}]{brin1998anatomy}
Brin, S. and Page, L. (1998).
\newblock The anatomy of a large-scale hypertextual web search engine.
\newblock {\em Computer networks and ISDN systems}, 30(1-7):107--117.

\bibitem[\protect\astroncite{Broder}{2002}]{broder2002taxonomy}
Broder, A. (2002).
\newblock A taxonomy of web search.
\newblock {\em ACM {SIGIR} Forum}, 36(2):3--10.

\bibitem[\protect\astroncite{Chen et~al.}{2011}]{chen-2011-speak}
Chen, J., Nairn, R., and Chi, E.~H. (2011).
\newblock Speak little and well: Recommending conversations in online social
  systems.
\newblock In {\em Proc CHI'11 Vancouver, BC, Canada}, pages 217--226.

\bibitem[\protect\astroncite{Chu et~al.}{2010}]{chu2010tweeting}
Chu, Z., Gianvecchio, S., Wang, H., and Jajodia, S. (2010).
\newblock Who is tweeting on twitter: human, bot, or cyborg?
\newblock In {\em Proceedings of the 26th annual computer security applications
  conference}, pages 21--30. ACM.

\bibitem[\protect\astroncite{Conover et~al.}{2011}]{cono-poli}
Conover, M.~D., Ratkiewicz, J., Francisco, M., Gon{\c c}alves, B., Flammini,
  A., and Menczer, F. (2011).
\newblock Political polarization on twitter.
\newblock In {\em Proc. AAAI ICWSM 5th Intl. Conf. Weblogs and Social Media
  2011}.

\bibitem[\protect\astroncite{Covington et~al.}{2016}]{covington2016deep}
Covington, P., Adams, J., and Sargin, E. (2016).
\newblock Deep neural networks for youtube recommendations.
\newblock In {\em Proceedings of the 10th ACM Conference on Recommender
  Systems}, pages 191--198. ACM.

\bibitem[\protect\astroncite{Datta et~al.}{2018}]{datta-2018-changing}
Datta, H., Knox, G., and Bronnenberg, B.~J. (2018).
\newblock Changing their tune: How consumers' adoption of online streaming
  affects music consumption and discovery.
\newblock {\em Marketing Science}, 37(1):5--21.

\bibitem[\protect\astroncite{Dillahunt et~al.}{2015}]{dillahunt-2015-detecting}
Dillahunt, T.~R., Brooks, C.~A., and Gulati, S. (2015).
\newblock Detecting and visualizing filter bubbles in google and bing.
\newblock In {\em Proc. CHI'15 Extended Abstracts, Apr 18-23, 2015, Seoul.}

\bibitem[\protect\astroncite{Dunbar et~al.}{2015}]{dunbar-2015-the-structure}
Dunbar, R., Arnaboldi, V., Conti, M., and Passarella, A. (2015).
\newblock The structure of online social networks mirrors those in the offline
  world.
\newblock {\em Social Networks}, 43:39--47.

\bibitem[\protect\astroncite{Dupret and Piwowarski}{2008}]{dupret2008user}
Dupret, G.~E. and Piwowarski, B. (2008).
\newblock A user browsing model to predict search engine click data from past
  observations.
\newblock In {\em Proceedings of the 31st annual international ACM SIGIR
  conference on Research and development in information retrieval}, pages
  331--338. ACM.

\bibitem[\protect\astroncite{Ekstrand et~al.}{2015}]{ekstrand-2015-letting}
Ekstrand, M.~D., Kluver, D., Harper, F.~M., and Konstan, J.~A. (2015).
\newblock Letting users choose recommender algorithms: An experimental study.
\newblock In {\em Proc. {ACM} {RecSys'15} Ninth ACM Conf. on Recommender
  Systems}, pages 11--18.

\bibitem[\protect\astroncite{Elberse}{2008}]{elberse-2008-should}
Elberse, A. (2008).
\newblock Should you invest in the long tail?
\newblock {\em Harvard Business Review}, 86(7/8):88--96.

\bibitem[\protect\astroncite{Elberse and
  Oberholzer-Gee}{2006}]{elberse-2006-superstars}
Elberse, A. and Oberholzer-Gee, F. (2006).
\newblock Superstars and underdogs: An examination of the long tail phenomenon
  in video sales.
\newblock Technical Report 07-015, Harvard Business School Working Paper.

\bibitem[\protect\astroncite{Epstein and Robertson}{2015}]{epst-sear}
Epstein, R. and Robertson, R.~E. (2015).
\newblock The search engine manipulation effect (seme) and its possible impact
  on the outcomes of elections.
\newblock {\em {PNAS}}, 112(33):E4512--E4521.

\bibitem[\protect\astroncite{Ferrara et~al.}{2016}]{ferr-rise}
Ferrara, E., Varol, O., Davis, C., Menczer, F., and Flammini, A. (2016).
\newblock The rise of social bots.
\newblock {\em Communications of the {ACM}}, 59(7):96--104.

\bibitem[\protect\astroncite{Fishbein and Ajzen}{2011}]{fishbein2011predicting}
Fishbein, M. and Ajzen, I. (2011).
\newblock {\em Predicting and changing behavior: The reasoned action approach}.
\newblock Psychology Press.

\bibitem[\protect\astroncite{Gl{\"o}ggler}{2013}]{gloggler2013onthepage}
Gl{\"o}ggler, M. (2013).
\newblock On the {P}age {M}ethoden der {O}ptimierung.
\newblock In {\em Suchmaschinen im Internet: Funktionsweisen, Ranking Methoden,
  Top Positionen}, chapter~6, pages 115--168. Springer-Verlag.

\bibitem[\protect\astroncite{Goel et~al.}{2010}]{goel-2010-anatomy}
Goel, S., Broder, A., Gabrilovich, E., and Pang, B. (2010).
\newblock Anatomy of the long tail: ordinary people with extraordinary tastes.
\newblock In {\em Proc. {WSDM'10} {ACM} 3rd Intl Conf on Web Search and Data
  Mining}, pages 201--210, New York, NY. ACM.

\bibitem[\protect\astroncite{Hargittai}{2000}]{hargittai-2000-open}
Hargittai, E. (2000).
\newblock Open portals or closed gates? channeling content on the world wide
  web.
\newblock {\em Poetics}, 27:233--253.

\bibitem[\protect\astroncite{Hindman et~al.}{2003}]{hind-goog}
Hindman, M., Tsioutsiouliklis, K., and Johnson, J.~A. (2003).
\newblock Googlearchy: How a few heavily-linked sites dominate politics on the
  web.
\newblock In {\em Annual Meeting of the Midwest Political Science Association},
  pages 1--33.

\bibitem[\protect\astroncite{Introna and
  Nissembaum}{2000}]{introna-2000-shaping}
Introna, L.~D. and Nissembaum, H. (2000).
\newblock Shaping the web: Why the politics of search engines matters.
\newblock {\em The Information Society}, 16:169--185.

\bibitem[\protect\astroncite{Joachims
  et~al.}{2007}]{Joachims:2007:EAI:1229179.1229181}
Joachims, T., Granka, L., Pan, B., Hembrooke, H., Radlinski, F., and Gay, G.
  (2007).
\newblock Evaluating the accuracy of implicit feedback from clicks and query
  reformulations in web search.
\newblock {\em ACM Trans. Inf. Syst.}, 25(2).

\bibitem[\protect\astroncite{Kitchin}{2017}]{kitchin2017thinking}
Kitchin, R. (2017).
\newblock Thinking critically about and researching algorithms.
\newblock {\em Information, Communication \& Society}, 20(1):14--29.

\bibitem[\protect\astroncite{Lee et~al.}{2011}]{lee2011seven}
Lee, K., Eoff, B.~D., and Caverlee, J. (2011).
\newblock Seven months with the devils: A long-term study of content polluters
  on twitter.
\newblock In {\em Proc. {AAAI} {ICWSM} 5th Intl. Conf. Weblogs and Social Media
  2011}, pages 185--192.

\bibitem[\protect\astroncite{Lessig}{2009}]{lessig2009code}
Lessig, L. (2009).
\newblock {\em Code: And other laws of cyberspace}.
\newblock ReadHowYouWant. com.

\bibitem[\protect\astroncite{Lewis et~al.}{2012}]{lewis-2012-social}
Lewis, K., Gonzalez, M., and Kaufman, J. (2012).
\newblock Social selection and peer influence in an online social network.
\newblock {\em {PNAS}}, 109(1):68--72.

\bibitem[\protect\astroncite{Liben-Nowell and Kleinberg}{2003}]{libe-link}
Liben-Nowell, D. and Kleinberg, J. (2003).
\newblock The link prediction problem for social networks.
\newblock In {\em CIKM '03: Proceedings of the 12th international conference on
  Information and knowledge management}, pages 556--559, New York, NY, USA. ACM
  Press.

\bibitem[\protect\astroncite{Licklider and Taylor}{1968}]{lick-comp}
Licklider, J. and Taylor, R. (1968).
\newblock The computer as a communication device.
\newblock {\em Science and technology}.

\bibitem[\protect\astroncite{Lietz et~al.}{2014}]{liet-when}
Lietz, H., Wagner, C., Bleier, A., and Strohmaier, M. (2014).
\newblock When politicians talk: Assessing online conversational practices of
  political parties on twitter.
\newblock In {\em Proc. AAAI ICWSM 8th Intl. Conf. Weblogs and Social Media},
  pages 285--294, Palo Alto, CA. AAAI Press.

\bibitem[\protect\astroncite{Marres}{2017}]{marres2017digital}
Marres, N. (2017).
\newblock {\em Digital sociology: The reinvention of social research}.
\newblock John Wiley \& Sons.

\bibitem[\protect\astroncite{Messing and
  Westwood}{2014}]{messing-2014-selective}
Messing, S. and Westwood, S.~J. (2014).
\newblock Selective exposure in the age of social media: Endorsements trump
  partisan source affiliation when selecting news online.
\newblock {\em Communication Research}, 41(8):1042--1063.

\bibitem[\protect\astroncite{Mitra et~al.}{2014}]{mitra-2014-on-user}
Mitra, B., Shokouhi, M., Radlinski, F., and Hofmann, K. (2014).
\newblock On user interactions with query auto-completion.
\newblock In {\em Proc. {ACM} {SIGIR} 37th Intl. Conf. on Research \&
  development in information retrieval}, pages 1055--1058. {ACM}.

\bibitem[\protect\astroncite{Mueller-Birn et~al.}{2013}]{mull-work}
Mueller-Birn, C., Dobusch, L., and Herbsleb, J. (2013).
\newblock Work-to-rule: the emergence of algorithmic governance in wikipedia.
\newblock In {\em Proc. 6th International Conference on Communities and
  Technologies (C\&T '13)}, pages 80--89, New York. ACM.

\bibitem[\protect\astroncite{Munson et~al.}{2013}]{munson-2013-encouraging}
Munson, S.~A., Lee, S.~Y., and Resnick, P. (2013).
\newblock Encouraging reading of diverse political viewpoints with a browser
  widget.
\newblock In {\em Proc. ICWSM 7th {AAAI} Intl. Conf. Weblogs and Social Media},
  pages 419--428. AAAI press.

\bibitem[\protect\astroncite{Munson and Resnick}{2010}]{munson-2010-presenting}
Munson, S.~A. and Resnick, P. (2010).
\newblock Presenting diverse political opinions: How and how much.
\newblock In {\em Proc. {CHI} 2010: Expressing and Understanding Opinions in
  Social Media, April 10--15, 2010, Atlanta, GA, USA}, pages 1455--1466.

\bibitem[\protect\astroncite{Murakami et~al.}{2010}]{murakami-2010-statement}
Murakami, K., Nichols, E., Mizuno, J., Watanabe, Y., Masuda, S., Goto, H.,
  Ohki, M., Sao, C., Matsuyoshi, S., Inui, K., and Matsumoto, Y. (2010).
\newblock Statement map: Reducing web information credibility noise through
  opinion classification.
\newblock In {\em Proc. of the 4th workshop on Analytics for noisy unstructured
  text data}, {AND '10}, pages 59--66, New York, NY. {ACM}.

\bibitem[\protect\astroncite{Neff and Nagy}{2016}]{neff-2016-talking}
Neff, G. and Nagy, P. (2016).
\newblock Talking to bots: Symbiotic agency and the case of tay.
\newblock {\em International Journal of Communication}, 10:4915--4931.

\bibitem[\protect\astroncite{Nguyen et~al.}{2014}]{nguyen2014exploring}
Nguyen, T.~T., Hui, P.-M., Harper, F.~M., Terveen, L., and Konstan, J.~A.
  (2014).
\newblock Exploring the filter bubble: the effect of using recommender systems
  on content diversity.
\newblock In {\em Proceedings of the 23rd international conference on World
  wide web}, pages 677--686. ACM.

\bibitem[\protect\astroncite{Niederer and van Dijck}{2010}]{nied-wisd}
Niederer, S. and van Dijck, J. (2010).
\newblock Wisdom of the crowd or technicity of content? {W}ikipedia as a
  sociotechnical system.
\newblock {\em New Media \& Society}, 12(8):1368--1387.

\bibitem[\protect\astroncite{Pariser}{2011}]{pari-filt}
Pariser, E. (2011).
\newblock {\em The Filter Bubble. What the Internet is Hiding from You}.
\newblock The Penguin Press, New York.

\bibitem[\protect\astroncite{Park et~al.}{2009}]{park-2009-newscube}
Park, S., Kang, S., Chung, S., and Song, J. (2009).
\newblock Newscube: delivering multiple aspects of news to mitigate media bias.
\newblock In {\em Proceedings of the SIGCHI Conference on Human Factors in
  Computing Systems}, CHI '09, pages 443--452, New York, NY. {ACM}.

\bibitem[\protect\astroncite{Rader and Gray}{2015}]{rader-2015-understanding}
Rader, E. and Gray, R. (2015).
\newblock Understanding user beliefs about algorithmic curation in the facebook
  news feed.
\newblock In {\em Proc. ACM CHI'15}, pages 173--182.

\bibitem[\protect\astroncite{Resnick et~al.}{2013}]{resnick-2013-bursting}
Resnick, P., Garrett, R.~K., Kriplean, T., Munson, S.~A., and Stroud, N.~J.
  (2013).
\newblock Bursting your (filter) bubble: Strategies for promoting diverse
  exposure.
\newblock In {\em CSCW '13 Companion, Feb. 23--27, 2013, San Antonio, Texas,
  USA}, pages 95--100.

\bibitem[\protect\astroncite{Salganik et~al.}{2006}]{salg-expe}
Salganik, M.~J., Dodds, P.~S., and Watts, D.~J. (2006).
\newblock Experimental study of inequality and unpredictability in an
  artificial cultural market.
\newblock {\em Science}, 311:854--856.

\bibitem[\protect\astroncite{Shah et~al.}{2013}]{shah-2013-exploration}
Shah, C., Liu, J., Gonz{\'a}lez-Ib{\'a}{\~n}ez, R., and Belkin, N. (2013).
\newblock Exploration of dynamic query suggestions and dynamic search results
  for their effects on search behaviors.
\newblock {\em Proceedings of the American Society for Information Science and
  Technology}, 49(1):1--10.

\bibitem[\protect\astroncite{Shao et~al.}{2018}]{shao2018spread}
Shao, C., Ciampaglia, G.~L., Varol, O., Yang, K.-C., Flammini, A., and Menczer,
  F. (2018).
\newblock The spread of low-credibility content by social bots.
\newblock {\em Nature Communications}, 9(1):4787.

\bibitem[\protect\astroncite{Shirky}{2006}]{shirky2006power}
Shirky, C. (2006).
\newblock Power laws, weblogs, and inequality.
\newblock In Dean, J., Anderson, J., and Lovink, G., editors, {\em Reformatting
  politics: Information Technology and Global Civil Society}, pages 35--42.
  Routledge, New York, NY.

\bibitem[\protect\astroncite{Siles}{2011}]{siles-2011-online}
Siles, I. (2011).
\newblock From online filter to web format: Articulating materiality and
  meaning in the early history of blogs.
\newblock {\em Social Studies of Science}, 41(5):737--758.

\bibitem[\protect\astroncite{Steck}{2011}]{steck-2011-item}
Steck, H. (2011).
\newblock Item popularity and recommendation accuracy.
\newblock In {\em Proc. RecSys'11, Oct 23-27, 2011, Chicago, IL}, pages
  125--132.

\bibitem[\protect\astroncite{Steiner}{2014}]{Steiner-014}
Steiner, T. (2014).
\newblock Bots vs. wikipedians, anons vs. logged-ins.
\newblock In {\em Proceedings of the 23rd International Conference on World
  Wide Web}, WWW '14 Companion, pages 547--548, New York, NY, USA. ACM.

\bibitem[\protect\astroncite{Tan et~al.}{2007}]{tan-2007-term}
Tan, B., Velivelli, A., Fang, H., and Zhai, C. (2007).
\newblock Term feedback for information retrieval with language models.
\newblock In {\em Proc. {ACM} {SIGIR}'07 Proceedings of the 30th international
  ACM SIGIR conference on Research and development in Information Retrieval}.
  {ACM}.

\bibitem[\protect\astroncite{Thomas et~al.}{2012}]{thomas-2012-adapting}
Thomas, K., Grier, C., and Paxson, V. (2012).
\newblock Adapting social spam infrastructure for political censorship.
\newblock In {\em Presented as part of the 5th {USENIX} Workshop on Large-Scale
  Exploits and Emergent Threats}, San Jose, CA. {USENIX}.

\bibitem[\protect\astroncite{Zhang and Nasraoui}{2006}]{zhang-2006-mining}
Zhang, Z. and Nasraoui, O. (2006).
\newblock Mining search engine query logs for query recommendation.
\newblock In {\em Proc. {ACM} WWW'06 15th Intl. Conf. World Wide Web}. {ACM}.

\bibitem[\protect\astroncite{Zhou et~al.}{2010}]{zhou2010impact}
Zhou, R., Khemmarat, S., and Gao, L. (2010).
\newblock The impact of {YouTube} recommendation system on video views.
\newblock In {\em Proceedings of the 10th ACM SIGCOMM conference on Internet
  measurement}, pages 404--410. ACM.

\end{thebibliography}
\end{document}